\documentstyle[12pt,myown,twoside,axodraw]{report}
\begin{document}
\protect{\pagenumbering{arabic}}
\protect{\setcounter{page}{1}}
\begin{center}
{\Large{\bf How Is Nature Asymmetric ?}}\\ \vspace{0.5 cm}
{\large{\sf 2. $CP$ and $T$ Violation in Elementary Particle Physics}}\\
\vspace{0.5 cm}
B. Ananthanarayan, J. Meeraa, Bharti Sharma, \\
Seema Sharma and Ritesh K. Singh\\
\end{center}
{\bf This two-part article considers certain
fundamental symmetries of nature, namely the discrete symmetries of parity (P),
charge conjugation (C) and time reversal (T), and their possible
violation. Recent experimental results are discussed in some depth.
In this second part, we discussion $CP$ and $T$ violation and arrive
at a synthesis.}\\\\
In this second part we discuss the violation of two discrete symmetries,
$CP$ and $T$, which follows our discussion in general of the discrete symmetries
$C$, $P$ and $T$, and that of parity violation in the first part.
We conclude with a synthesis of the ideas presented in the two parts of
this article.\\\\
{\large{\bf CP-invariance}}\\\\
In the weak interactions, $P$ and $C$ are maximally violated simultaneously,
such that the system is symmetric under the combined operation of $CP$. This is
clear from the way massless fermions transform under $P$ and $C$ operations
(Fig. 1).\\\\
In nature only left-handed neutrino ($\nu_L$) and right-handed anti-neutrino
($\bar\nu_R$) exist, which are $CP$ transforms of each other. Hence, earlier it
was thought that though weak interactions violate $P$ and $C$ they still
possess $CP$-symmetry.\\\\
In 1964, Christenson,  Cronin, Fitch and Turlay observed that decay of neutral
kaon violates $CP$-symmetry minutely. Neutral kaons, $K^0$ and $\bar K^0$, are
produced in pion-proton collisions via strong interaction,
$$\pi^- p  \longrightarrow \ K^0 \Lambda \hspace{2cm}\pi^+ p \longrightarrow
\bar K^0 K^+ p$$
and they decay primarily to two-pion and three-pion final states via the weak
interaction with two different life-times. Based on the life-times of decays in
to these modes, kaons were renamed as $K$-short ($\tau=0.89\times 10^{-10}
sec$) and $K$-long ($\tau=5.17\times 10^{-8} sec$). By conservation of angular
momentum and intrinsic parity of pion,  it is clear that the two-pion state,
$|2 \ \pi\rangle$\footnote{It is common in quantum mechanics to represent a
state of a physical system by writing down certain of its characteristics
inside the symbol $| \ \rangle $, called {\em ket}.}, has $CP=+1$ and the
three-pion state, $|3 \ \pi\rangle$, has $CP=-1$. This implies that
$|K^0\rangle$ and
$|\bar K^0\rangle$ are not $CP$ eigenstates, and in fact they transform into
each other under $CP$ operation i.e.,
$$|K^0\rangle \stackrel{CP}{\longrightarrow} |\bar{K^0}\rangle$$
With this property of neutral kaon states the $CP$ eigenstates can be formed as
the following linear superpositions (see Box 1),
$$|K^0_1\rangle \ = \ \frac{1}{\sqrt{2}}\left(|K^0\rangle + |\bar
K^0\rangle\right) \hspace{2cm}CP=+1$$
$$|K^0_2\rangle \ = \ \frac{1}{\sqrt{2}}\left(|K^0\rangle - |\bar
K^0\rangle\right) \hspace{2cm}CP=-1$$
Earlier $K^0_1$ ($K^0_2$) was identified as $K_S$ ($K_L$), which decays to two
(three)-pion or $CP$-even (odd) state, and $CP$ is conserved. {\bf But it was
observed that $K_L$ also decays into two-pions, which is $CP$-even and hence
$CP$ is violated.}\\\\
If we take a beam of $K^0$, which is produced by strong interactions, and which
can be written as linear superposition of $K_L$ and $K_S$, then the two-pion
decay mode of $K_L$ will interfere with that of $K_S$ as a function of time.
The intensity of the pion beam in two-pion decay mode varies with time as,
\begin{eqnarray}
I_{2\pi}(t)&=&I_{2\pi}(0) \ \left[ \ e^{-\Gamma_S t} + |\eta_{+-}|^2
e^{-\Gamma_L t} \right.\nonumber\\
&+& \left.2|\eta_{+-}|  e^{-(\Gamma_S +\Gamma_L)t/2}\cos(\Delta m t + \phi_{+-
})\right]
\end{eqnarray}
where, $\Gamma_{L,S}$ = decay width of $K_{L,S}$, $\Delta m = m_L - m_S$, $t$
is time and,
$$\eta_{+-} \ = \ |\eta_{+-}| e^{i \phi_{+-}} \ = \ \frac{{\cal M}(K_L
\rightarrow \pi^+\pi^-)}{{\cal M}(K_S \rightarrow \pi^+\pi^-)}$$
where, the ${\cal M}$s stand for the {\it matrix elements} signifying the
transition probability amplitudes.\\\\
This interference was observed by Christenson {\it et. al.} in 1964. Similar to
$\eta_{+-}$, we can define another quantity, $\eta_{00}$, which is ratio of
amplitudes of decay in to $\pi^0\pi^0$ instead of $\pi^+\pi^-$. Nonvanishing
of the $\eta$'s implies that $CP$ is violated. Further $\eta$'s are written as,
$$\eta_{+-} \ = \ \epsilon + \epsilon' \hspace{1cm} \eta_{00} \ = \  \epsilon -
2\epsilon'$$
and $CP$ violation is generally expressed in terms of $\epsilon$ and
$\epsilon'$. Experimental values of these parameters are :
$$\epsilon \ = \ (2.271 \pm 0.017) \times 10^{-3}$$
$$\epsilon'/\epsilon \ = \ (2.1 \pm 0.5) \times 10^{-3}$$
A non-zero value of $\epsilon'$ is referred to as direct $CP$ violation. In the
Standard Model (SM) (see Box 2) the quantity $\epsilon'$ is somewhat smaller as
compared to the experimental values. But due to large calculational and
experimental uncertainties, the SM predictions are not ruled out. Other place
to see $CP$ violation is the $B$-meson system, which is similar to the $K$-
meson system, with the $s$-quark replaced by a $b$-quark. In SM the source of
$CP$ violation is Cabibbo Kobayashi Maskawa matrix (Box 2). The $CP$ violation
effect in $K$ must translate to $B$-mesons also.  At several particle physics
laboratories, ``asymmetric colliders" have been constructed to study the $B^0 -
\bar B^0$ system (see Box 3).\\\\
Further, according to $CPT$-theorem, $CP$ violation implies $T$ violation, but
there is no direct evidence for it in neutral kaon system. Thus the source of
$CP$ violation is not yet clear and the phenomenon needs better understanding
as it may have been responsible for the net {\it baryon number} of the present
day Universe (Box 4).\\\\
{\large{\bf T reversal violation}}\\\\
Soon after the discovery of parity violation in
experiments following the suggestions of Yang and Lee, many authors
considered the possibility of violation of the other discrete symmetries
we have discussed earlier.  The great Russian physicist L. D. Landau
considered the possibility of elementary particles possessing a
non-vanishing electric dipole moment (e.d.m.) which would imply the violation
ofT-reversal invariance.  Note however that a complex system like a water
molecule does possess an e.d.m., but this does not come into conflict with
T-reversal invariance.  There are many atomic systems in which the e.d.m.s
of elementary particles such as neutrons or electrons can manifest
themselves.  However, these e.d.m.s appear to be very small quantities and
there has been no detection of such effects.  This implies that
experiments must seek higher levels of precision before they can announce
a discovery, and it also implies that theoretical scenarios which predict
large values of e.d.m.s can be constrained or ruled out by the
non-observation of e.d.m.s. (The Standard Model of the
electroweak interaction gives a contribution to the neutron e.d.m. of the
order of $10^{-31}$ to $10^{-33}$ e cm which, because it is
second order in the weak
interaction coupling constant, is very small.)
Here we discuss one specific technique based on so-called
ultra-cold neutrons, which is implemented at Rutherford Appleton
Laboratories.
Nuclear reactors serve as copious sources of neutrons which are
used for a variety of experiments.  Normally these neutrons emerge
with a kinetic energy of about 1/40 eV, and are called thermal neutrons.
In order to carry out very precise measurements of static and other
properties of neutrons, it is necessary to slow them down to very low
kinetic energies of the order $10^{-7}$ eV or even less.  Such neutrons are
called ultra-cold neutrons and provide an opportunity to carry out highly
precise experiments.\\\\
``The Ramsey resonance technique'' can be used to measure with very
high precision the precession frequency of ultracold neutrons in a weak
magnetic field.  The precession frequency will change in the presence of
an electric field if the neutron has an e.d.m.  The most recent result
give an upper bound of the order of $d_n \le 10^{-26}$ e cm.\\\\
{\large{\bf Synthesis}}\\\\
In this two part article we have discussed the origins of the important
discrete symmetries, $C$, $P$, $T$ and $CP$, which are at the heart
of the nature of space and time. In particular, the fundamental
interactions between elementary  particles and the mediators of forces
between them must either respect these symmetries or violate them in a
specific manner, but without coming in conflict with the $CPT$ theorem, a
rigorous consequence of the special theory of relativity.
 Of special interest are current and
ongoing experiments that verify and constrain theories of these
interactions at an ever increasing level of precision. Among these is
the violation of $P$ by the weak-interactions, which has 
recently been detected in a
low-energy but high precision experiment involving the nuclear anapole
moment.  The violation of $CP$ in the $B$-meson system is only the second
example of such a phenomenon discovered in current experiments.  The
detection of direct $T$ violation is an ongoing experimental challenge in
such settings as the ultra-cold neutron facilities.  $CP$ violation is also
crucial for the rise of baryon asymmetry from a baryon symmetric Universe
in the standard big-bang cosmology.  In this brief article, we have
examined and summarized all the aspects mentioned above of a fascinating
and challenging aspect of the physical world in which we live.\\\\
{\large{\bf Acknowledgements}}\\\\
JM, BS and SS thank the summer students programme of
the Centre for Theoretical Studies, Indian Institute of Science which led
to this article. We thank Dr. B. P. Das for innumerable discussions, and Dr.
B. Moussallam for a careful reading of the manuscript.\\\\
{\bf Box 1 : Superposition principle of Quantum Mechanics}\\\\
One of the postulates of Quantum Mechanics (QM) which makes it different form
classical mechanics is the principle of {superposition}  of quantum states. The
state of the system is described by a {\it wave function}, say $\psi$, which
satisfies Schr\"odinger equation
$$i\hbar \frac{\partial}{\partial t}\psi(x,t)=-\frac{\hbar^2}{2m}
\bigtriangledown^2 \psi(x,t)+V(x) \psi(x,t)$$
where $V(x)$ is potential. The probability of finding the system between $x$
and $x+dx$ at time $t$ is $|\psi(x,t)|^2 dx$ and hence $\psi$ is also called
probability
amplitude. The superposition principle states that {\it under logical OR
operation
its the amplitudes which adds and not the probabilities}. For example, if two
possible solutions of Schr\"odinger equation are $\psi_1$ and $\psi_2$ then
probability
of finding the system between $x$ and $x+dx$ at time $t$ in either of the two
states is $|\psi_1(x,t)+\psi_2(x,t)|^2 dx$, and not
$(|\psi_1(x,t)|^2+|\psi_2(x,t)|^2) dx$ as
expected classically.\\\\
The concept of superposition in QM has an analogy in optics. If $L_x$ and $L_y$
denote the amplitudes of plane polarised light polarised along $x$ and $y$
directions
respectively, then $L_{\pm}=(L_x\pm iL_y)/\sqrt{2}$ denote the amplitudes of
right
and left circularly polarised light. We know that light (or photons) has only
two polarisation states, and both \{$L_x$, $L_y$\} and \{$L_+$, $L_-$\} are
complete and equivalent description of photon polarisation. We may choose
either of them depending upon the symmetry of the system we are dealing with.
In the same spirit if a system
is described by a set of  $n$ different states denoted by \{$\psi_i, \ i=1,
2,...,n$\}, then any other set \{$\phi_i, \ i=1,2,...,n$\} describes the system
equivalently if $\phi_i$'s are linearly independent when written as linear
superposition of $\psi_i$'s.\\\\
{\bf Box 2:  The Standard Model of Electroweak and Strong interactions}\\\\
Based on decades of experiments in the field of elementary particle
physics, today a picture which is referred to as the Standard Model of
interactions has come into being.  It describes the interactions between
several types of particles, which come in two varieties.  The first
variety are called leptons, which participate only in the electroweak
interactions and not in the strong interactions, and the second variety
are called hadrons which participate in both.  For example, the electron
and its neutrino are leptons.  They have heavier cousins called the muons
and $\tau$ leptons and their respective neutrinos.  The hadrons themselves
come in two classes, namely mesons and baryons.  Among mesons we find the
pions, kaons and the B-mesons, whereas the proton and neutron are examples
of baryons.  These are understood in the SM to be arising from the
interactions of further constituents called quarks, which themselves come
in six ``flavors,'' the up, down, strange, charm, bottom and top quarks.
These particles typically carry electric charge as do the charged leptons
and also additional quantum numbers called colors through which they
interact amongst themselves with the exchange of so-called gluons which
mediate the strong interactions.  The electroweak forces themselves are
mediated by the well-known photon and other particles called $W^{\pm}$ and $Z$.
\\\\
The SM, as proposed by Glashow, Weinberg and Salam, had only first four flavors
of quarks and possesses $CP$ symmetry. To explain $CP$ violation in $K$-meson
system two more quarks namely $b$ and $t$ were introduced. In SM, the source of
CP violation is mixing between quarks which is described by Cabbibo-Kobayashi-
Maskawa (CKM) matrix. The matrix elements of the relevent $2\times 2$ sub-
matrix are real if we have only four flavors of quarks and this implies that
$CP$ is a good symmetry. $CP$ will be violated only if some of the CKM matrix
elements are complex, which is possible when we have at least six quarks in
nature (details are beyond the scope of this article). The present day SM
contains six quarks and can accomodate $CP$ violation in $K$ and $B$ meson
systems.\\\\
{\bf Box 3: The ``Asymmetric" B-factory.}\\\\
Conventional cyclotron design of elementary particle colliders consists of
a ring in which a beam of particles of a certain type, say electrons
(protons), is accelerated in one sense, while the anti-particles,
say positrons (anti-protons) are accelerated in the opposite sense in the
{\it same} ring, which is guaranteed by the fact that the particle and
anti-particle differ only in the sign of their electric charge, while
their masses are equal.  Many successful experiments for generations
were based on this design, the latest include the Large Electron Positron
collider at CERN, which will be upgraded to the Large Hadron Collider.
 However, the B-factories are based on a novel
design which was forced upon by the limitation of technology which
required that the B-mesons should travel a distance larger than the spatial
resolution
of the silicon vertex detectors that are used.  The asymmetric design uses
one of the fundamental implications of Einstein's special theory of
relativity, which is that of time-dilation as follows:  the electron beam
energy is significantly larger than that of the positron beam, each of
which is now accelerated in different rings and then brought together to a
central detector region.  The asymmetry of the energies ensures that the
decay products are now boosted in the laboratory frame and therefore the
particles generated ``live longer" and traverse a distance that is large
enough for the detectors to resolve.   As a concrete example, we consider
the BELLE experiment at Japan's KEK laboratory.  Here the electron energy
corresponds to $E_e= 8 GeV$, (where the $GeV$ unit corresponds roughly to
the energy that would be generated if a proton were to be converted into
energy), while the positron energy corresponds to $E_{\overline{e}}= 3.5
GeV$.  This corresponds to $\beta=v/c=0.391$, which is the boost velocity of
the reaction products in the laboratory frame in units of the velocity of light
and $\gamma=1/\sqrt{1-\beta^2}=1.18$, the {\it time dilation} factor (or the
reciprocal of the{\it Lorentz-Fitzgerald} contraction factor).  The life-time
of the
$B$-meson is $1.55\times 10^{-12}$ sec, which implies that in a symmetric
collider it would have
only traversed 35 $\mu m$, while in BELLE it traverses 290 $\mu m$, while the
silicon vertex detector resolution corresponds to 50 $\mu m$.  We thus
illustrate with this example the interplay of the special theory of
relativity, the needs of elementary particle physics research and present
day technology.\\\\
{\bf Box 4: CP violation and the early Universe}\\\\
The currently accepted picture of the origin of the Universe is
often referred to as the Big Bang picture.  The Big Bang out of
which the known Universe is supposed to have arisen is based on
solutions of Einstein's equations of Genreral Relativity which correspond to an
expanding, homogeneous and isotropic Universe, and which accounts for
the recession of galaxies governed by Hubble's law, and for the
nearly homogeneous cosmic microwave background radiation corresponding
to a black body temperature of 2.7 K.  In our Universe, we observe
more matter than anti-matter (``baryon asymmetry").  One may pose a
question as to whether in the Big Bang epoch this was the case.  The
well-known Russian physicist Andrei Sakharov answered this question as
follows:  a baryon symmetric Universe at the time of the big bang could have
evolved into an asymmetric one if three conditions were necessarily met.
\begin{itemize}
\item there must be interactions present which violate baryon number in the
first place,\item that the Universe be out of thermal equilibrium, and
\item that CP be violated.
\end{itemize}
In this manner CP violation could possibly hold
one of the keys to the origin of the matter-antimatter asymmetry in the
Universe
as we know it.  In fact,
one of the reasons cited for the award of the Nobel prize to Fitch and
Cronin is the connection of CP violation to this feature of the Universe.\\\\
{\bf Suggested Readings}\\\\
A very inspiring book on the subject of symmetries is the following
classic:
\begin{itemize}
\item H. Weyl, {\it Symmetry}, Princeton University Press, Princeton, NJ, USA,
1952
\end{itemize}
Standard introductory reference to nuclear and elementary particle
physics are, e.g.,
\begin{itemize}
\item B. Povh, K. Rith, C. Scholz and F. Zetsch, 2nd edn.,
{\it Particles and Nuclei, An Introduction to the Physical Concepts},
Springer-Verlag, Berlin, 1999.
\item D. Griffiths, {\it Introduction to Elementary Particle Physics},
John Wiley \& Sons, New York, 1987
\item  G.D. Coughlan, J.E. Dodd {\it The Ideas of Particle Physics : An
Introduction for Scientists},Cambridge Univ. Pr., UK, 1993
\end{itemize}
The following articles in Resonance would provide the reader with
accounts on some of the subjects disccused here:
\begin{itemize}
\item Ashoke Sen, {\it Resonance}, Vol.5, No.1, p.4 2000
\item Rohini Godbole, {\it Resonance}, Vol.5, No.2, p.16 2000
\item Sourendu Gupta, {\it Resonance}, Vol.6, No.2, p.29 2001
\end{itemize}
A detailed treatment of the Cobalt experiment has already been made
available to the readers of Resonance in:
\begin{itemize}
\item Amit Roy, {\it Resonance}, Vol.6, No.8, p.32 2001
\end{itemize}
A readily accessible treatment of the subject of T violation maybe found
in book
\begin{itemize}
\item R. G. Sachs, {\it The Physics of Time Reversal},
University of Chicago Press, Chicago, 1987
\end{itemize}
The subject of CP violation has recently been discussed in two very
detailed books:
\begin{itemize}
\item G. C. Branco, L. Louvra and J. P. Silva, {\it CP Violation},
Oxford University Press, New York, 1999
\item I. I. Bigi and A. Sanda, {\it CP Violation}, Cambridge University Press,
Cambridge, UK,2000
\end{itemize}
It is possible to trace recent developments in this field on the world
wide web using appropriate keywords and search engines.  We do not list
the URLs here.
\newpage
.\\
\vspace{5cm}
\begin{figure}[h]
\begin{center}
\begin{picture}(120,120)(-10,-10)
\Line(30,5)(70,5)       \Line(30,-5)(70,-5)
\Line(30,5)(30,10)      \Line(30,-5)(30,-10)
\Line(30,10)(20,0)      \Line(30,-10)(20,0)
\Line(70,5)(70,10)      \Line(70,-5)(70,-10)
\Line(70,10)(80,0)      \Line(70,-10)(80,0)
{\SetOffset(0,100)
\Line(30,5)(70,5)       \Line(30,-5)(70,-5)
\Line(30,5)(30,10)      \Line(30,-5)(30,-10)
\Line(30,10)(20,0)      \Line(30,-10)(20,0)
\Line(70,5)(70,10)      \Line(70,-5)(70,-10)
\Line(70,10)(80,0)      \Line(70,-10)(80,0)
}
\Line(5,30)(5,70)       \Line(-5,30)(-5,70)
\Line(5,30)(10,30)      \Line(-5,30)(-10,30)
\Line(10,30)(0,20)      \Line(-10,30)(0,20)
\Line(5,70)(10,70)      \Line(-5,70)(-10,70)
\Line(10,70)(0,80)      \Line(-10,70)(0,80)
{\SetOffset(100,0)
\Line(5,30)(5,70)       \Line(-5,30)(-5,70)
\Line(5,30)(10,30)      \Line(-5,30)(-10,30)
\Line(10,30)(0,20)      \Line(-10,30)(0,20)
\Line(5,70)(10,70)      \Line(-5,70)(-10,70)
\Line(10,70)(0,80)      \Line(-10,70)(0,80)
} 
\Line(30,70)(30,56)     \Line(30,70)(44,70)
\Line(70,30)(56,30)     \Line(70,30)(70,44)
\Line(30,56)(33.5,59.5) \Line(44,70)(40.5,66.5)
\Line(56,30)(59.5,33.5) \Line(70,44)(66.5,40.5)
\Line(33.5,59.5)(59.5,33.5)     \Line(40.5,66.5)(66.5,40.5)
\Text(0,0)[]{\Large $\bar\nu_L$}
\Text(100,0)[]{\Large $\bar\nu_R$}
\Text(0,100)[]{\Large $\nu_L$}
\Text(100,100)[]{\Large $\nu_R$}
\Text(50,115)[]{\Large $P$}     \Text(50,-15)[]{\Large $P$}
\Text(115,50)[]{\Large $C$}     \Text(-15,50)[]{\Large $C$}
\Text(65,65)[]{\Large $CP$}
\end{picture}
\end{center}
\caption{\small{Transformation of massless fermions under $P$ and $C$ and their
combined operation.}}
\end{figure}
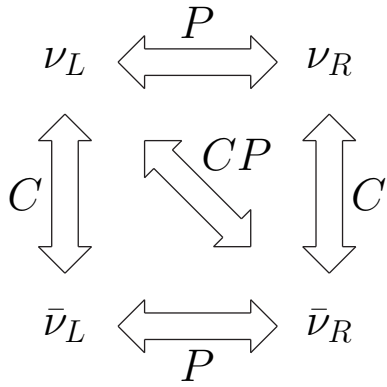
\end{document}